\documentclass[sigplan,10pt]{acmart}

\renewcommand\footnotetextcopyrightpermission[1]{}
\setcopyright{none}

\usepackage[USenglish]{babel}     
\usepackage{xcolor}               
\usepackage{amsmath,amsfonts}     
\usepackage{booktabs}             
\usepackage{array,multirow,tabularx,makecell,threeparttable}
\usepackage{enumitem}
\usepackage{xspace}
\usepackage{textcomp}
\usepackage{pifont}
\usepackage{url}
\usepackage{subcaption}           
\usepackage{listings}             
\usepackage[ruled,linesnumbered]{algorithm2e}
\SetArgSty{textnormal}
\usepackage{tikz}
\usepackage{balance}
\usepackage{soul}
\usepackage{relsize}
\usepackage{comment}
\usepackage{configures/fancyhdr}  

\newcolumntype{L}[1]{>{\raggedright\arraybackslash}p{#1}}
\newcolumntype{C}[1]{>{\centering\arraybackslash}p{#1}}
\newcolumntype{R}[1]{>{\raggedleft\arraybackslash}p{#1}}
\newcolumntype{T}{>{\hsize=2.5\hsize\linewidth=2.5\hsize}X}

\newcommand{\SystemName}{\textsc{CrossPool}\xspace}

\begin{document}

\title[]{\SystemName: Efficient Multi-LLM Serving for Cold MoE Models through KV-Cache and Weight Disaggregation}
\author{
    Zhuoren Ye,
    Tianyu Wo,
    Dinghao Xue,
    Mingming Zhang,
    Yuchen Teng,
    Chunming Hu,
    Renyu Yang\textsuperscript{\dag}
}
\thanks{\textsuperscript{\dag}Corresponding Author}
\affiliation{
    School of Software, Beihang University
    \country{}
}
\email{{yezr, woty, dinghaoxue, mingmingzhang, yuchenteng, hucm, renyuyang}@buaa.edu.cn}
\renewcommand{\authors}{Zhuoren Ye, Tianyu Wo, Dinghao Xue, Mingming Zhang, Yuchen Teng, Chunming Hu, Renyu Yang}
\renewcommand{\shortauthors}{Z. Ye, T. Wo, D. Xue et al.}

\begin{abstract}

Emerging LLM services increasingly host many sparse MoE models, yet most models
receive sparse requests and remain cold. This creates a
GPU memory problem: model weights are stable and model-determined, while KV-cache
is transient and demand-determined. Because cold models rarely reach peak KV-cache
demand at the same time, reserving worst-case KV capacity per model wastes memory;
a shared KV-cache pool can instead provision aggregate active demand. However, KV-cache sharing is not sufficient when weights and KV-cache remain in a
monolithic GPU memory pool. Static weights compete with dynamic KV-cache, and
KV-head-limited attention under cold, low-concurrency traffic exposes only a
fraction of replicated KV capacity, leading to low GPU memory utilization and weak
long-context support. We present \SystemName, a serving engine for
cold MoE models that separates FFN weights and KV-cache into two GPU memory pools:
a weights pool that consolidates FFN weights across cold models, and a KV-cache
pool that dynamically serves active requests while keeping attention local to
KV-cache. \SystemName combines a KV-cache planner and virtualizer, a layer-wise
pipeline scheduler that hides hidden-state transfers, and persistent kernels with
control lowering to reduce CPU-GPU control overhead. With efficient GPU memory pooling,
\SystemName underpins bursty long-context requests and outperforms the state-of-the-art \texttt{kvcached}-based multi-LLM serving system, reducing P99
TBT by up to $10.4\times$.

\end{abstract}

\maketitle

\section{Introduction}

Emerging large language models (LLMs) now power applications ranging from
chatbots~\cite{chatgpt, gemini, deepseek} to agentic
assistants~\cite{claudecode, codex, openclaw}. To serve this demand, model
providers~\cite{deepinfra, alibabamodelstudio, volcanoengine} deploy
increasingly capable, fast-evolving LLMs in their datacenters. Many recent
frontier and open-weight models support long context for complex tasks and use
Mixture-of-Experts (MoE) architectures~\cite{
DBLP:journals/corr/abs-2401-04088,
DBLP:conf/acl/DaiDZXGCLZYWXLH24,
DBLP:journals/corr/abs-2505-09388,
DBLP:journals/corr/abs-2412-19437,
DBLP:conf/iclr/LepikhinLXCFHKS21,
DBLP:journals/jmlr/FedusZS22,
deepseek-v4} to scale total parameters while keeping computation moderate.

This trend makes cold-model serving increasingly important. Recent
studies~\cite{DBLP:conf/icml/DuanLDLZLS024, DBLP:conf/usenix/Gao0ZLS25,
DBLP:conf/sosp/Xiang0QYZYZL0025} show that a few \textit{hot models} serve
most requests, while many \textit{cold models} are underutilized. A serving
cluster that keeps many such models online therefore pays the memory cost of
large MoE weights even when most models have few active requests. Long-context
serving adds a second pressure point: KV-cache can grow to a large fraction of
GPU memory, but unlike weights, it is allocated only while requests are active and is
reclaimed after decode finishes.

The key opportunity is that weights and KV-cache have different lifetimes.
Model weights are \textit{model-determined} and \textit{stable}, whereas
KV-cache is \textit{demand-determined} and \textit{transient}: its footprint
varies with request rate, context length, and concurrency. Since cold models
rarely peak simultaneously, reserving worst-case KV-cache for each model wastes
GPU memory. Hence, a shared KV-cache pool can instead provision for aggregate active
demand.

Simple memory sharing within a unified GPU memory pool is insufficient. Existing
multi-LLM serving systems~\cite{DBLP:conf/icml/DuanLDLZLS024, yu2026chimera}
improve GPU sharing through multiplexing or elastic memory management, but they
still colocate static weights and dynamic KV-cache under the same GPU pool.
This couples the KV-cache capacity visible to a request with the amount of
memory already occupied by weights. It also exposes an \textit{algorithm-system
mismatch}: KV-head-limited attention algorithms such as
MLA~\cite{DBLP:journals/corr/abs-2405-04434} and
MQA~\cite{DBLP:journals/corr/abs-1911-02150, deepseek-v4} reduce per-token
KV-cache, while serving engines often use DP attention for such models to
increase aggregate throughput. Under cold, low-concurrency workloads, however,
only replicas holding active requests expose usable KV-cache capacity to those
requests. The result is not merely inefficient sharing: modern long-context MoE
models can lose usable context capacity under the same hardware budget.

We present \SystemName, an LLM serving framework for cold MoE models that
separates FFN weights and KV-cache into different GPU memory pools. The weights
pool consolidates FFN weights from multiple cold models, while the KV-cache pool
dynamically shares GPU memory for active KV-cache across model instances.
\SystemName keeps attention and other non-FFN modules in the KV-cache pool for
local KV access, and executes FFN modules in the weights pool because they
dominate MoE parameters. The pool boundary exchanges hidden states rather than
KV-cache tensors. This design increases shared KV-cache capacity without giving
up competitive performance. To make this practical, \SystemName addresses
three design challenges: heterogeneous KV-cache planning, communication across
disaggregated pools, and graph/control overhead under layer-wise mixed scheduling.
Our main technical contributions are threefold.
\begin{itemize}[leftmargin=0.8em]
    \item \textit{KV-cache planner and virtualizer}. \SystemName plans the
    shared KV-cache pool budget and parallelism offline, then exposes the pool
    through virtualized paging~\cite{nvidia-vmm}.

    \item \textit{Layer-wise pipeline scheduler}. \SystemName schedules
    attention and FFN at layer granularity, overlapping hidden-state transfers
    with computation across in-flight batches.

    \item \textit{Persistent kernels and control lowering}. \SystemName
    separately captures attention and FFN subgraphs and lowers frequent
    scheduling and communication control to GPU-resident persistent kernels.
\end{itemize}
We evaluate \SystemName on a production-grade platform against Static Partition
and \texttt{kvcached}~\cite{yu2026chimera}. \SystemName improves memory pooling
efficiency to support long-context requests while preserving competitive
performance on balanced workloads.

\begin{figure}[t]
    \centering
    \begin{subfigure}[b]{0.45\columnwidth}
        \includegraphics[width=\linewidth]{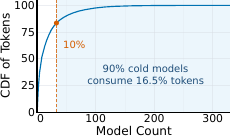}
        \caption{Underutilized cold models}
        \label{fig:underutilized-cold-models}
    \end{subfigure}
    \begin{subfigure}[b]{0.45\columnwidth}
        \includegraphics[width=\linewidth]{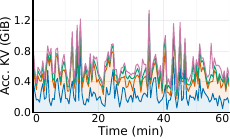}
        \caption{Accumulated KV-cache}
        \label{fig:acc-kv}
    \end{subfigure}
    \vspace{-0.3em}
    \caption{
        Cold-model underutilization and accumulated KV-cache usage under low RPS. Subfigure
        (a) summarizes data from OpenRouter~\cite{openrouter}, and (b) stacks active KV-cache bytes
        for four 7B models at 0.2 RPS over one hour.
    }
\end{figure}

\section{Background and Motivation}
\label{sec:background-and-motivation}

\subsection{Multi-LLM Serving and KV-cache Memory}
\label{sec:multi-serving-and-kv-memory}

LLM model providers (\textit{e.g.,} DeepInfra~\cite{deepinfra}, Alibaba Model
Studio~\cite{alibabamodelstudio}, Volcano Engine~\cite{volcanoengine}) are
deploying more LLMs to meet rising demand for agentic applications. Recent
work~\cite{DBLP:conf/icml/DuanLDLZLS024, DBLP:conf/usenix/Gao0ZLS25,
DBLP:conf/sosp/Xiang0QYZYZL0025} shows that LLM popularity is highly skewed:
a few \textit{hot models} receive most user requests, while many \textit{cold
models} are underutilized. Fig.~\ref{fig:underutilized-cold-models} plots the
CDF of token consumption across models on OpenRouter~\cite{openrouter} and shows
that $\sim$90\% of models are cold and consume few tokens.

Serving cold models is intricate because of their large GPU memory footprint.
During inference, GPU memory is mainly occupied by model weights and KV-cache.
Weights are fixed by the architecture and remain resident once a model is
loaded, while KV-cache is generated dynamically by attention and reclaimed
after requests finish. Fig.~\ref{fig:acc-kv} shows that at low RPS (requests per
second), active KV-cache usage varies widely across models but often remains
low, indicating a potential to share a unified KV-cache pool across multiple
cold models. It is desirable to properly size a shared pool for aggregate
active KV-cache at a random time, not for each request's peak. Under low request
volume (\textit{cold traffic}), many models exhibit no active requests, and
KV-cache usage fluctuates as requests start and finish. This allows
provisioning the pool to a high percentile of aggregate demand instead of the
worst-case load for each model.

\begin{figure}[t]
    \centering
    \begin{subfigure}[b]{\linewidth}
        \centering
        \includegraphics[width=0.9\linewidth]{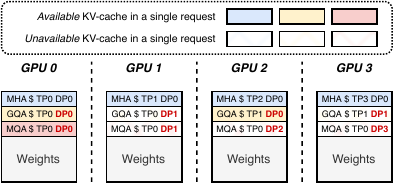}
        \caption{Aggregated memory pool}
        \label{fig:comparison-a}
        \vspace{1em}
    \end{subfigure}
    \begin{subfigure}[b]{\linewidth}
        \centering
        \includegraphics[width=0.9\linewidth]{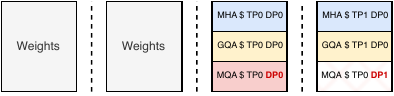}
        \caption{Disaggregated memory pool} 
        \label{fig:comparison-b}
    \end{subfigure}
    \vspace{-0.4em}
    \caption{
        KV-cache availability when serving a single request on 4 GPUs. Comparison of
        monolithic and disaggregated memory pools for weights and KV-cache.
        \texttt{n\_heads} values of MHA, GQA and MQA are 4, 2 and 1, respectively.
    }
    \label{fig:comparison}
\end{figure}

\subsection{Mismatch between Algorithms and Systems}\label{sec:mismatch}
Recent LLMs use diverse attention algorithms (\textit{e.g.,}
GQA~\cite{DBLP:conf/emnlp/AinslieLJZLS23}, MQA~\cite{DBLP:journals/corr/abs-1911-02150},
MLA~\cite{DBLP:journals/corr/abs-2405-04434})
to improve computation and memory efficiency. For KV-cache placement, the key
factor is the number of KV heads. With $G$ GPUs, these algorithms fall 
into: (\textbf{Type I}) MHA and GQA with \texttt{n\_heads} $\ge G$; and
(\textbf{Type II}) MQA, MLA, and GQA with \texttt{n\_heads} $< G$. TP
(\textit{a.k.a.,} tensor/head parallelism) naturally exposes
KV-cache capacity for Type I models.

Modern serving engines (\textit{e.g.,} vLLM~\cite{DBLP:conf/sosp/KwonLZ0ZY0ZS23} and SGLang~\cite{DBLP:conf/nips/ZhengYXS0YCKSGB24}) typically use data parallelism (\textit{a.k.a.,} DP) or DP attention (\textit{a.k.a.,} DPA) for KV-head-limited models. These replicas offer high aggregate KV-cache capacity under high concurrency, but in cold-model serving only replicas with active requests provide usable capacity. Existing multi-LLM serving engines inherit these parallelism schemes from their backends and therefore suffer from  \textit{low effective KV-cache utilization} for Type II algorithms in low-concurrent cases.

Furthermore, current multi-LLM serving engines (e.g.,
MuxServe~\cite{DBLP:conf/icml/DuanLDLZLS024} and
\texttt{kvcached}~\cite{yu2026chimera}) typically adopt
placement strategies that colocate model weights and KV-cache within a single GPU
memory pool. As a result, the cumulative footprint of model weights occupies a substantial fraction of the limited GPU memory, thereby 
yielding \textit{low per-GPU KV-cache capacity}. This, in turn, diminishes support for long-context workloads and can, in extreme cases, lead to out-of-memory (OOM) failures or request rejection.

Fig. \ref{fig:comparison-a} illustrates how current multi-LLM serving engines allocate all
model parameters and KV-cache in a monolithic GPU memory pool. Under a single-request workload,
in the MHA configuration (\texttt{n\_heads} = 4, implying TP = 4), the KV-cache is effectively distributed across all four
GPUs, and the request can utilize the aggregate KV-cache capacity of the entire 4-GPU group. In contrast, with grouped-query attention (GQA) (\texttt{n\_heads} = 2, hence TP = 2 and DP = 2), an individual request is confined to a single two-GPU
replica, and therefore can only access the KV-cache capacity of that replica. Under multi-query attention (MQA) (\texttt{n\_heads} = 1, hence DP = 4), each
 request is restricted to the KV-cache of a single GPU in this placement scheme. Consequently, the 1/2 and 1/4
 factors represent the fraction of total KV-cache capacity that is visible to a single request under low-concurrency conditions. In addition, the
colocation of model parameters and KV-cache within the same GPU memory further reduces the effective KV-cache capacity available on each GPU.

\subsection{Challenges}
\label{sec:challenges}

As shown in Fig. \ref{fig:comparison-b}, it is desirable to decouple the
monolithic GPU memory pool into two disaggregated pools -- a model weights pool
and a KV-cache pool. This enables flexible attention parallelism, improves
KV-cache utilization by avoiding unnecessary DP replicas, and maximizes per-GPU
KV-cache capacity. However, this organization must solve three challenges.

\noindent\textbf{C1: Enabling diverse KV-cache strategies.}
Different attention algorithms produce different KV tensor layouts, numbers of
KV heads, and per-token KV bytes. Therefore, a shared KV-cache pool cannot use a
fixed pool budget or parallelism strategy without accounting for the colocated
models. The system must estimate aggregate KV demand from prompt length, output
length, service time, and arrival rate; account for model-specific KV bytes per
token; and choose parallelism plans that expose usable KV-cache capacity under
cold concurrency. The output of this planning must be enforceable online as a
pool page budget and rank assignments, otherwise the shared pool can be
over-admitted during a burst.

\noindent\textbf{C2: Non-negligible communication overhead between pools.}
Disaggregating weights and KV-cache creates a new pool boundary in the inference
path. Attention must access KV-cache locally because the KV tensors grow with
context length and are read at every decode step. Moving KV-cache tensors across
the boundary would therefore make attention communication-bound, especially for
long-context requests. At the same time, not all weights should stay with the
KV-cache pool, because MoE FFN modules dominate the stored parameters. The pool
boundary must instead exchange hidden states, whose size is bounded by batch
tokens and hidden dimension rather than accumulated context length. However,
hidden-state transfer is still on the critical path: it occurs at every
pool crossing, for every layer, and for every generated token. Even though each
transfer is much smaller than moving KV-cache tensors, the repeated transfers
accumulate into non-negligible communication overhead.

\noindent\textbf{C3: Increased graph capture complexity under mixed scheduling.}
Modern serving engines rely on CUDA graph capture to reduce launch overhead, but
disaggregated execution introduces layer-level breakpoints between attention and
FFN. Serving multiple cold models across separate pools further creates mixed
scheduling: different models can have different layer counts, batch sizes, and
active request timing. If each layer transition and communication step is driven
from the host, cold batches suffer frequent CPU-GPU control transitions and the
benefit of graph capture is reduced. The system must retain graph execution for
attention and FFN subgraphs while lowering frequent scheduling and communication
control off the host-side path.

\vspace{-1em}
\section{\SystemName Design}
\label{sec:design}

\begin{figure}[t]
    \centering
    \includegraphics[width=0.85\linewidth]{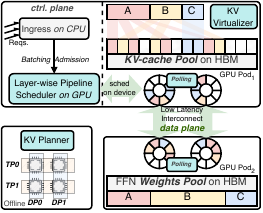}
    \caption{\SystemName system architecture}
    \label{fig:sys-arch}
\end{figure}

We propose \SystemName, a multi-LLM serving system that colocates cold MoE
models by separately pooling GPU memory for model weights and KV-cache
(Fig.~\ref{fig:sys-arch}). The weights pool aggregates weights from multiple
cold models, and the KV-cache pool provides on-demand KV-cache space for active
requests. During decode, a request batch first executes attention and other
non-FFN modules in the KV-cache pool, transfers the resulting hidden states to
the weights pool for FFN execution, and transfers the FFN output back to the
KV-cache pool for the next layer.

\SystemName keeps attention close to KV-cache to avoid transferring KV-cache between
two pools. It places non-FFN modules in the KV-cache pool for local KV
access and executes FFN modules in the weights pool. This partition matches MoE
parameter structure, where FFN modules hold most weights
(Table~\ref{tab:weight-breakdown}). It also defines a communication boundary
whose traffic does not grow with context length.

\begin{table}[t]
    \centering
    \caption{Weight breakdown for five 10 -- 30B models.}
    \label{tab:weight-breakdown}
    \scriptsize
    \begin{threeparttable}
    \resizebox{\columnwidth}{!}{%
    \begin{tabular}{@{}lcrrrrr@{}}
        \toprule
        Model & Type & Layers & FFN & Attn. & FFN share \\
        \midrule
        DeepSeek-V2-Lite & MoE & 27 & 14.9B & 0.4B & 95.0\% \\
        Qwen3-30B-A3B & MoE & 48 & 29.0B & 0.9B & 95.0\% \\
        GLM-4.7-Flash & MoE & 47 & 28.3B & 1.0B & 94.5\% \\
        Llama-30B & Dense & 60 & 21.5B & 10.6B & 66.0\% \\
        Qwen3-32B & Dense & 64 & 25.2B & 6.0B & 76.8\% \\
        \bottomrule
    \end{tabular}}
    \end{threeparttable}
\end{table}

The two pools communicate over low-latency links; our prototype uses NVLink
with NVSHMEM~\cite{nvshmem} for hidden-state transfer. The following
subsections describe how KV-cache planner and virtualizer work in response to
\textbf{C1}, the layer-wise pipeline scheduler for \textbf{C2} and persistent
kernels with control lowering for \textbf{C3}.

\subsection{KV-cache Planner and Virtualizer}\label{sec:virt-kv}
\SystemName includes an offline KV-cache planner and an online KV-cache
virtualizer for heterogeneous KV-cache management. The planner computes the
shared KV-cache pool budget and the parallelism plan for each model. The
virtualizer then presents the pool through a uniform paged KV-cache interface,
so attention kernels can use a logical KV address space while the virtualizer
controls physical page allocation.

The planner uses per-model prompt, output, and service-time samples, arrival
rate $\lambda_M$, page size, per-token KV bytes $\kappa(M)$, and a target
provisioning quantile. It emits two outputs: the total KV-cache pool page
budget and a parallelism plan that determines how each model uses the KV-cache
pool. For model $M$, request $i$ arrives at $A_i$ and has
prompt tokens $O_{p,i}$, output tokens $O_{d,i}$, and decode residence time
$T_i$ in the KV pool. The planner provisions for aggregate active KV at a random
observation time, rather than the final KV footprint of every request. At
request age $u$, active KV tokens grow as
\begin{equation}
    \begin{gathered}
        Q_i(u)= \left(O_{p,i}+O_{d,i}\frac{u}{T_i}\right)\mathbf{1}_{\{0\le u<T_i\}},\\
        K_M(t)=\sum_i\kappa(M)Q_i(t-A_i).
    \end{gathered}
    \label{equ:kv-active}
\end{equation}
The aggregate pool demand is:
\begin{equation}
    K_{\mathrm{pool}}(t)=\sum_MK_M(t).
\end{equation}
\SystemName uses the P95/P99 trace-driven Monte Carlo
quantile~\cite{dong2018tutorial} as a conservative provisioning target. This
sampling keeps empirical correlations among prompt length, output length, and
service time, which are lost if the planner independently sizes each dimension
by a worst-case percentile. After the pool demand is estimated, the planner
rounds memory to pages and sets the shared KV-cache pool page budget.

During online serving, \SystemName schedules each request batch to the DP rank
that has the largest free KV-cache space among ranks eligible for the model. The
virtualizer employs CUDA VMM APIs~\cite{nvidia-vmm} to reserve a virtual KV
address range for each model and map physical KV pages on demand. Attention
operators see a normal paged KV-cache interface, while physical page mapping and
unmapping remain slow paths outside the per-token critical path.
Active KV pages are kept until their decode requests finish; if the pool page
budget is exhausted, admission control queues or rejects new requests instead
of interrupting active decode requests.

\begin{figure}[t]
    \centering
    \includegraphics[width=0.9\linewidth]{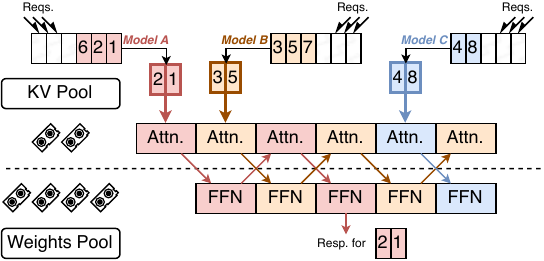}
    \caption{
        Layer-wise pipeline scheduler. It interleaves attention and FFN layers of two
        batches, allowing attention and FFN to be executed simultaneously on different
        batches from different models. Early exit is supported when one batch finishes
        all its layers.
    }
    \label{fig:pipeline}
\end{figure}

\subsection{Layer-wise Pipeline Scheduler}\label{sec:pipeline-sched}
To address \textbf{C2}, \SystemName uses a layer-wise pipeline scheduler to reduce
exposed ping-pong overhead between attention in the KV-cache pool and FFN in the
weights pool by overlapping their execution across batches.
Fig. \ref{fig:pipeline} shows the design. The scheduler maintains
two in-flight batches, each with its own model id, layer cursor, and completion
state. When batch $B_1$ executes attention for a layer in the KV-cache pool, the
hidden states of batch $B_2$ from a previous layer can be processed by FFN kernels
in the weights pool. After an attention stage finishes, the scheduler launches
hidden-state transfer to the weights pool; after an FFN stage finishes, it
transfers the output back to the KV-cache pool and advances that batch's layer
cursor.

This layer-granular state machine keeps both pools busy without requiring the
two batches to have the same model or the same number of layers. If one batch
finishes all its layers, the scheduler publishes its tokens, releases the
completed batch state, and fetches a new batch from the corresponding request
queue. The other batch can continue from its current layer cursor. This avoids a
global layer barrier across models and makes the pipeline naturally compatible
with heterogeneous cold-model colocation.

\begin{figure}
    \centering
    \begin{subfigure}[t]{\linewidth}
        \centering
        \includegraphics[width=\linewidth]{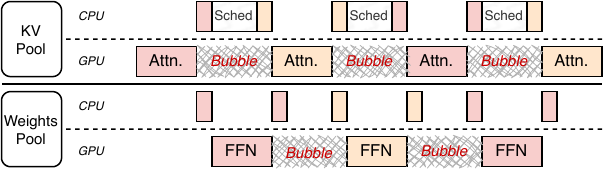}
        \caption{CPU-launched kernels and communication}
        \label{fig:persistent-kernel-a}
        \vspace{1em}
    \end{subfigure}
    \begin{subfigure}[t]{\linewidth}
        \centering
        \includegraphics[width=\linewidth]{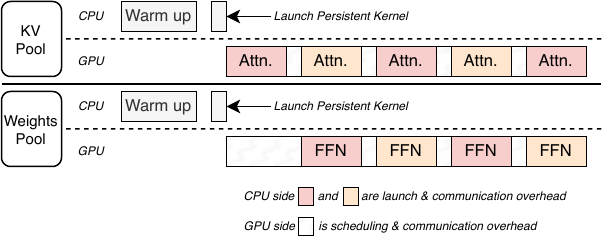}
        \caption{Persistent kernels with control lowering}
        \label{fig:persistent-kernel-b}
    \end{subfigure}
    \vspace{-1em}
    \caption{
        Persistent kernels for efficient graph execution.
    }
\end{figure}

\subsection{Persistent Kernels and Control Lowering}
\label{sec:persistent-kernel}

To tackle \textbf{C3}, \SystemName captures graphs for attention and FFN separately and
triggers JIT compilation during warmup. Separating the two graph types preserves
graph execution within each pool while still allowing layer-wise scheduling
across the pool boundary. \SystemName then uses persistent kernels and control
lowering to keep frequent scheduling and communication control on GPUs,
reducing host intervention and CPU-GPU control transitions. Fig.
\ref{fig:persistent-kernel-b} illustrates the design.

\SystemName captures supported attention and FFN subgraphs during warmup and
passes their graph handles to GPU-resident persistent kernels. These kernels are
launched at the beginning of serving and keep running persistently. Each
persistent kernel repeatedly polls per-model request queues on GPUs, selects the
next ready batch for its pool, dispatches the corresponding captured graph,
launches inter-pool hidden-state communication when the graph reaches a pool
boundary, waits for computation and communication completion, and updates the
batch state for the next layer. This design keeps the high-frequency control
loop on GPUs, while the host remains responsible for coarse-grained admission,
KV-cache page mapping, and failure handling.

\section{Implementation}
\label{sec:implementation}

We implement \SystemName based on SGLang v0.5.10. The implementation extends the
runtime with disaggregated memory pools, a transport layer for hidden
state exchange, virtualized KV-cache paging in the KV-cache pool, and persistent
GPU-side dispatch for layer-wise execution.

\noindent\textbf{Runtime integration.}
\SystemName keeps the SGLang request scheduler, tokenizer path, sampling path,
and attention kernels in the KV-cache pool. The main runtime change is at the
transformer layer boundary. In the KV-cache pool, FFN layers are replaced with
proxy layers. A proxy layer receives the hidden states produced by the local
attention and non-FFN modules, sends them to the weights pool, waits for the
remote FFN output, and then resumes the next layer in the KV-cache pool. The
weights pool hosts the corresponding FFN modules and executes them using SGLang
FusedMoE kernels. This structure keeps most existing SGLang operator
implementations intact and model adaptation reusable while moving the dominant
MoE weights out of the KV-cache pool.

\noindent\textbf{Inter-pool transport.}
We add a transport library based on NVSHMEM~\cite{nvshmem} for low-latency
hidden-state transfer between pools. The transport interface is placed exactly
at the proxy-FFN boundary: the KV-cache pool sends hidden states to the weights
pool, and the weights pool sends FFN outputs back. Communication is issued
asynchronously so that the layer-wise scheduler can overlap attention and FFN
computation from different batches.

\noindent\textbf{KV-cache virtualization.}
The KV-cache pool uses CUDA VMM APIs~\cite{nvidia-vmm} to expose virtualized
KV-cache space to attention operators. The runtime reserves virtual address
ranges according to the planner output and maps physical pages as the allocator
slow path. Existing SGLang attention kernels (backed by
FlashAttention~\cite{DBLP:conf/nips/DaoFERR22},
FlashInfer~\cite{DBLP:conf/mlsys/00010LLZW0KGKC25}, etc.) continue to operate
over a paged KV-cache abstraction, while \SystemName tracks physical page
allocation against the shared pool page budget. Active decode pages are kept
until their requests finish; when the planned pool budget cannot admit a new
request, admission control queues or rejects it.

\noindent\textbf{Graph capture and persistent dispatch.}
\SystemName captures attention and FFN subgraphs separately during warmup and
passes their graph handles to persistent kernels. Our current implementation
captures batch sizes 1--4, which are typical for cold-model decode serving and
cover the optimized path used in our experiments. The persistent kernels poll
GPU-resident queues, dispatch captured graphs, launch inter-pool communication,
and update batch state without returning to the host for every layer transition.

\noindent\textbf{End-to-end integration.}
\SystemName focuses on decode-side memory colocation. Following
Aegaeon~\cite{DBLP:conf/sosp/Xiang0QYZYZL0025}, prefill phase runs on separate
temporal-multiplexing engines. Decode phase uses the disaggregated KV-cache and
weights pools, so our evaluation reports decode-side scalability, TBT, and
output throughput.

\section{Experiments}
\label{evaluation}

\subsection{Experimental Setup}

\noindent\textbf{Workloads and models}.
We use Vicuna ShareGPT~\cite{sharegpt-vicuna, sharegpt-vicuna-dataset}
conversations as a balanced input/output workload for the decode-side TBT experiment and
LongAlign~\cite{DBLP:conf/emnlp/BaiLZHQH0DL24, longalign10k}
as a long-context workload for the scalability experiment.
Request arrivals follow a Poisson process with low RPS and token counts are
calculated with each model's tokenizer. We evaluate baselines and \SystemName on
three MoE models colocation:
Qwen3-\allowbreak{}30B-\allowbreak{}A3B,
GLM-\allowbreak{}4.7-\allowbreak{}Flash, and
DeepSeek-\allowbreak{}V2-\allowbreak{}Lite. Accommodating weights of these 3
models requires $\sim$154 GB HBM.

\noindent\textbf{Baselines}.
We compare \SystemName against (1) \textit{Static Partition}: SGLang native static placement,
which assigns each model a fixed GPU placement and uses static MIG partitioning when
feasible; and (2) \textit{Chimera}~\cite{yu2026chimera} (\textit{i.e.,} \texttt{kvcached}): elastic
GPU memory pooling with \texttt{kvcached} which dynamically manages KV memory across models, but does
not disaggregate KV-cache and FFN weights into separate pools. We run both of the baselines
and \SystemName in prefill-decode disaggregation mode, and focus the comparison on the
decode phase.

\noindent\textbf{Testbeds and placement}.
Our experiments use a five-A100 server (40 GB HBM per GPU with NVLINK) for
decode-phase evaluation. Table~\ref{tab:eval-placement} summarizes the GPU
placement used by each system. These placements use the minimum GPU count
needed to serve the three models and keep the same total GPU memory budget
for model weights and KV-cache across systems. For Static Partition and
\texttt{kvcached}, we enable DP attention for GLM-4.7-Flash and
DeepSeek-V2-Lite, which adopt MLA, to avoid KV-cache replication.

\noindent\textbf{Metrics}.
Because \SystemName primarily optimizes decode-side memory colocation, we report
two metrics. First, we report the maximum aggregate request rate that a system
can sustain as context length increases. For the scalability experiment, we
estimate max RPS by sampling LongAlign context lengths and applying each
system's GPU placement and KV-cache budget; the estimate corresponds to the
offered rate before out-of-memory errors or fatal serving failures. Second, we
report TBT (Time-Between-Tokens) as the primary decode latency metric. We finally
report aggregate output throughput in an ablation study to isolate the impact
of two mechanisms that reduce disaggregation overhead.

\begin{figure}[t]
    \centering
    \includegraphics[width=\linewidth]{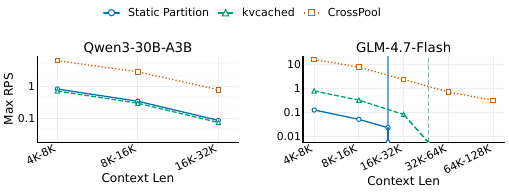}
    \vspace{-0.4em}
    \caption{
        Maximum aggregate request rate estimated from sampled LongAlign
        context-length bins within each model's nominal context window; vertical
        drops mark per-system capacity limits. Prompts are truncated to maximum
        context length of the two models.
    }
    \vspace{-1em}
    \label{fig:eval-max-rps}
\end{figure}

\begin{table}[t]
    \centering
    \caption{GPU placement on the five-A100 testbed.}
    \label{tab:eval-placement}
    \scriptsize
    \begin{threeparttable}
    \resizebox{\columnwidth}{!}{%
    \begin{tabular}{@{}lccccc@{}}
        \toprule
        Solution & GPU 0 & GPU 1 & GPU 2 & GPU 3 & GPU 4 \\
        \midrule
        Static Partition & Q & Q & G & G & D \\
        \texttt{kvcached} & Q + D & Q + G & Q + G & Q + G & G + D \\
        \SystemName & KV & W & W & W & W \\
        \bottomrule
    \end{tabular}}
    \begin{tablenotes}[flushleft]
        \scriptsize
        \item Q = Qwen3-30B-A3B, G = GLM-4.7-Flash, D = DeepSeek-V2-Lite
        \item KV = KV-cache, and W = FFN weights of Q + G + D.
    \end{tablenotes}
    \end{threeparttable}
\end{table}

\subsection{Context-length Scalability}\label{subsec:ctx-scalability}

Fig.~\ref{fig:eval-max-rps} evaluates whether \SystemName can serve bursty
long-context requests under the same five-GPU decode environment. As context
length grows, each active request consumes more KV-cache, so a fixed per-model
memory partition eventually reaches a capacity cliff. We sample LongAlign
requests within each non-empty long-context bin and estimate max RPS according
to the GPU placement and KV-cache budget used by Static Partition, \texttt{kvcached}, and
\SystemName. For DP-attention configurations, the figure reports aggregate RPS summed
across replicas; DP does not increase the maximum context length that any single
replica can admit.

The vertical drops in Fig.~\ref{fig:eval-max-rps} show where each placement no
longer has enough KV-cache to admit the sampled context lengths. Static
Partition and \texttt{kvcached} lose the ability to serve longer LongAlign bins once the
per-model or per-replica KV capacity is exhausted. In contrast, \SystemName keeps
a positive max RPS throughout the plotted supported range by exposing a larger
shared KV-cache pool under the same hardware budget. This result demonstrates
the capability \SystemName targets for cold serving: handling bursty
long-context traffic without giving each model a large dedicated KV-cache
partition.

\subsection{Overall Performance}

\begin{figure}[t]
    \centering
    \includegraphics[width=\linewidth]{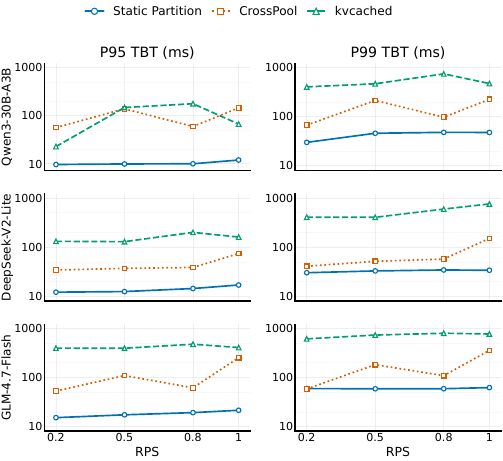}
    \vspace{-0.4em}
    \caption{
        Decode-side TBT on ShareGPT traces from 0.2 to 1.0 RPS per model.
    }
    \vspace{-1em}
    \label{fig:eval-tbt}
\end{figure}

According to the context-length scalability analysis in §\ref{subsec:ctx-scalability},
and our LongAlign serving runs, Static Partition and
\texttt{kvcached} can encounter OOM or fatal serving failures on long-context
requests even at low request rates. We therefore use Vicuna ShareGPT as a balanced
input/output workload for the TBT comparison, where all systems can complete the
same request-rate sweep. Fig.~\ref{fig:eval-tbt} compares the
ShareGPT TBT performance of Static Partition, \texttt{kvcached},
and \SystemName. We sweep the request rate from 0.2 to 1.0 RPS and report P95
and P99 TBT for the three evaluated models. This experiment evaluates whether
\SystemName preserves decode latency while colocating the three cold MoE models
through shared KV-cache and weight pools.

Across the ShareGPT sweep, \SystemName remains competitive with
\texttt{kvcached} while providing the pooling capacity shown in
Fig.~\ref{fig:eval-max-rps}. At
0.8 RPS, \SystemName reduces P99 TBT over \texttt{kvcached} by $7.6\times$ for
Qwen3-30B-A3B, $10.4\times$ for DeepSeek-V2-Lite, and $7.3\times$ for
GLM-4.7-Flash. At 1.0 RPS, the corresponding P99 reductions are $2.1\times$,
$5.0\times$, and $2.2\times$.

Kvcached shows higher tail TBT mainly because elastic colocation allows
uncoordinated multi-model concurrency, which leads to SM, memory-bandwidth, and
KV-cache contention during decode. Its shared weight/KV memory pool and limited
per-replica KV capacity under DP attention further increase memory pressure under
bursty traffic. \SystemName separates the KV-cache pool from the FFN-weight pool
and uses a planned shared KV-cache budget, making resource usage more predictable.
Static Partition has the lowest latency because it keeps each model in a fixed
isolated placement, but this placement does not provide the long-context pooling
ability evaluated in Fig.~\ref{fig:eval-max-rps}. \SystemName instead targets
the middle ground: under the same hardware budget, it supports larger shared
KV-cache capacity while maintaining competitive token-generation latency.

\subsection{Ablation Study}

\begin{table}[t]
    \centering
    \caption{
        Ablation study of \SystemName techniques on ShareGPT, reporting
        aggregate decode-phase output throughput when serving three cold MoE
        models at 0.5 RPS per model.
    }
    \label{tab:eval-ablation}
    \scriptsize
    \resizebox{\columnwidth}{!}{%
    \begin{tabular}{@{}ccc@{}}
        \toprule
        \makecell{Layer-wise\\pipeline} &
        \makecell{Persistent kernels\\and control lowering} &
        \makecell{Aggregate decode\\output throughput (token/s)} \\
        \midrule
        Off & Off & 55.42 \\
        Off & \textbf{On} & 77.86 \\
        \textbf{On} & Off & 60.83 \\
        \hline
        \textbf{On} & \textbf{On} & \textbf{111.40} \\
        \bottomrule
    \end{tabular}}
    \vspace{-1em}
\end{table}

Table~\ref{tab:eval-ablation} isolates the throughput impact of the two
mechanisms (§\ref{sec:pipeline-sched} and §\ref{sec:persistent-kernel}) that reduce
disaggregation overhead. The first row disables both mechanisms and serves as the
unoptimized disaggregated baseline, where attention and FFN execution are serialized
across the pool boundary and layer-level control remains on the host-side path.
Enabling persistent kernels and control lowering alone improves aggregate decode
output throughput from 55.42 to 77.86 token/s, a $1.41\times$ gain, by reducing
frequent host-side dispatch and inter-pool communication control. Enabling the
layer-wise pipeline scheduler alone improves throughput to 60.83 token/s,
a $1.10\times$ gain, by overlapping attention and FFN execution across
in-flight batches.

Combining both mechanisms raises throughput to 111.40 token/s, a $2.01\times$
gain over the unoptimized disaggregated baseline. This shows that the two
mechanisms are complementary: persistent kernels reduce the control overhead on
the layer-wise path, while the scheduler keeps the attention and FFN pools busy
once that path can be driven efficiently.

\section{Discussion}
\label{sec:discussion}

\SystemName is designed for serving multiple cold MoE models, rather than
generic GPU sharing. Its main design choice is to separate two memory objects
with different lifetimes: FFN weights are stable and tied to the model, whereas
KV-cache is transient and tied to demand. \SystemName therefore consolidates the
dominant MoE FFN weights in the weights pool and keeps attention and other
non-FFN modules in the KV-cache pool, where attention can access KV-cache locally.
The pool boundary exchanges hidden states rather than KV-cache tensors, limiting
cross-pool traffic to data whose size is bounded by batch tokens and hidden
dimension. The layer-wise scheduler and persistent kernels then reduce the exposed
cost of this boundary by overlapping attention and FFN execution across in-flight
batches and keeping frequent dispatch and communication control on GPUs.

\noindent\textbf{Limitations and future directions.}
First, communication overhead is reduced but not fully hidden. The current
scheduler overlaps attention and FFN computation across batches, but A-to-F
(attention to FFN) hidden-state transfer from the KV-cache pool to the weights pool,
and F-to-A (FFN to attention) result transfer back to the KV-cache pool, remain
on the critical path. A future implementation could use a finer pipeline that
explicitly stages attention computation, A-to-F communication, FFN computation,
and F-to-A communication, so that communication stages can be overlapped with
adjacent compute stages more aggressively.

Second, heterogeneous model compute times can create pipeline imbalance.
Different MoE models can have different attention and FFN execution times per
layer. When a short-running model is pipelined with a long-running model, the
shorter stage can wait for the longer stage, aligning its observed decode time
with the slower model and reducing the benefit of colocation. This limitation can
be mitigated at cluster placement time by grouping cold models with similar
attention/FFN compute profiles before assigning them to the same disaggregated
pools.

\section{Related Work}

\noindent\textbf{Multi-LLM serving}.
MuxServe~\cite{DBLP:conf/icml/DuanLDLZLS024} and
Weaver~\cite{DBLP:conf/usenix/Gao0ZLS25} improve GPU sharing via spatial multiplexing.
Chimera~\cite{yu2026chimera} provides elastic memory management with
\texttt{kvcached}. Aegaeon~\cite{DBLP:conf/sosp/Xiang0QYZYZL0025} uses GPU
pooling and temporal multiplexing for skewed model popularity. These systems
motivate cold-model colocation, but they do not separate stable model weights
from transient KV-cache as two GPU memory pools. As a result, the KV-cache
capacity available to an active cold model is still constrained by the placement
and footprint of colocated weights. \SystemName instead disaggregates weights
and KV-cache to address the algorithm-system mismatch of KV-head-limited attention
under cold, low-concurrency workloads and thus improve shared KV-cache capacity for
requests.

\noindent\textbf{KV-cache management}.
vLLM~\cite{DBLP:conf/sosp/KwonLZ0ZY0ZS23} uses PagedAttention, and
SGLang~\cite{DBLP:conf/nips/ZhengYXS0YCKSGB24} uses RadixAttention, providing
efficient KV-cache management in a serving engine. Other systems expand,
compress, stream, or reuse KV-cache through external storage, host memory,
or selective prefetching, including Mooncake~\cite{DBLP:conf/fast/QinLHCRZ0ZX25},
LMCache~\cite{DBLP:journals/corr/abs-2510-09665},
CacheGen~\cite{DBLP:conf/sigcomm/LiuLCRHZDY0AMHH24}, and
InfiniGen~\cite{DBLP:conf/osdi/LeeLSS24}. These techniques reduce KV-cache
pressure for long-context serving, but they do not make GPU KV-cache a shared
resource across multiple cold MoE models. \SystemName keeps the primary inference
path on GPUs and uses virtualized paging to enforce a planned page budget for
the shared KV-cache pool.

\noindent\textbf{Attention-MoE disaggregation}.
MegaScale-Infer~\cite{DBLP:conf/sigcomm/ZhuJJWSWZZWCXZL25}
and Step-3~\cite{DBLP:journals/corr/abs-2507-19427} disaggregate attention and
FFN computation to scale MoE inference, improve throughput, and support
heterogeneous multi-node deployment. \SystemName uses a similar disaggregation
strategy, but aiming to enlarge shared KV-cache pool for multiple cold
MoE models via minimizing number of attention replicas. \SystemName typically
deploys KV-cache pool and weights pool in one or two nodes, utilizing low-latency
interconnection for cold-model serving.

\section{Conclusion}

Aggregate cold MoE serving exposes an algorithm-system mismatch. Model weights
are large and stable, while KV-cache is transient and unevenly used across cold models.
Existing multi-LLM systems improve GPU sharing but still store static weights
and dynamic KV-cache in one monolithic pool, which reduces both KV-cache usable
space and per-GPU capacity when serving recent KV-head-limited models, leading
to inefficient KV-cache GPU memory pooling and weak long-context support.

This paper presents \SystemName, a multi-LLM serving engine that splits FFN
weights and KV-cache into disaggregated GPU memory pools while keeping attention
local to KV-cache. \SystemName uses a trace-driven KV-cache planner and virtualized
paging to handle heterogeneous KV-cache demand, a layer-wise pipeline scheduler
to overlap attention and FFN across pools, and persistent kernels with control
lowering to cut host dispatch overhead. Our evaluation demonstrates context
scalability and competitive performance of \SystemName, comparing with
current state-of-the-art systems.

\bibliographystyle{ACM-Reference-Format}
\bibliography{references}

@inproceedings{DBLP:conf/sosp/Xiang0QYZYZL0025,
  author       = {Yuxing Xiang and
                  Xue Li and
                  Kun Qian and
                  Yufan Yang and
                  Diwen Zhu and
                  Wenyuan Yu and
                  Ennan Zhai and
                  Xuanzhe Liu and
                  Xin Jin and
                  Jingren Zhou},
  title        = {Aegaeon: Effective {GPU} Pooling for Concurrent {LLM} Serving on the
                  Market},
  booktitle    = {{SOSP}},
  pages        = {1030--1045},
  publisher    = {{ACM}},
  year         = {2025}
}

@inproceedings{DBLP:conf/sosp/KwonLZ0ZY0ZS23,
  author       = {Woosuk Kwon and
                  Zhuohan Li and
                  Siyuan Zhuang and
                  Ying Sheng and
                  Lianmin Zheng and
                  Cody Hao Yu and
                  Joseph Gonzalez and
                  Hao Zhang and
                  Ion Stoica},
  title        = {Efficient Memory Management for Large Language Model Serving with
                  PagedAttention},
  booktitle    = {{SOSP}},
  pages        = {611--626},
  publisher    = {{ACM}},
  year         = {2023}
}

@inproceedings{DBLP:conf/nips/ZhengYXS0YCKSGB24,
  author       = {Lianmin Zheng and
                  Liangsheng Yin and
                  Zhiqiang Xie and
                  Chuyue Sun and
                  Jeff Huang and
                  Cody Hao Yu and
                  Shiyi Cao and
                  Christos Kozyrakis and
                  Ion Stoica and
                  Joseph E. Gonzalez and
                  Clark W. Barrett and
                  Ying Sheng},
  title        = {SGLang: Efficient Execution of Structured Language Model Programs},
  booktitle    = {NeurIPS},
  year         = {2024}
}

@inproceedings{DBLP:conf/icml/DuanLDLZLS024,
  author       = {Jiangfei Duan and
                  Runyu Lu and
                  Haojie Duanmu and
                  Xiuhong Li and
                  Xingcheng Zhang and
                  Dahua Lin and
                  Ion Stoica and
                  Hao Zhang},
  title        = {MuxServe: Flexible Spatial-Temporal Multiplexing for Multiple {LLM}
                  Serving},
  booktitle    = {{ICML}},
  series       = {Proceedings of Machine Learning Research},
  pages        = {11905--11917},
  publisher    = {{PMLR} / OpenReview.net},
  year         = {2024}
}

@inproceedings{DBLP:conf/usenix/Gao0ZLS25,
  author       = {Shiwei Gao and
                  Qing Wang and
                  Shaoxun Zeng and
                  Youyou Lu and
                  Jiwu Shu},
  title        = {Weaver: Efficient Multi-LLM Serving with Attention Offloading},
  booktitle    = {{USENIX} {ATC}},
  pages        = {587--595},
  publisher    = {{USENIX} Association},
  year         = {2025}
}

@inproceedings{yu2026chimera,
  title        = {Chimera: Cost-Efficient Multi-LLM Serving via GPU Memory Ballooning},
  author       = {Yu, Shan and Qiao, Yifan and Ma, Mingyuan and Li, Yangmin and Yang, Shuo and Tong, Xinyuan and Wang, Yang and Xie, Zhiqiang and An, Yuwei and Cao, Shiyi and Bao, Ke and Vij, Deepak and Ding, Xiaoning and Wang, Yichen and Lu, Qingda and Wang, Zhong and Gao, Gao and Xu, Harry and Shu, Junyi and Xing, Jiarong and Sheng, Ying},
  booktitle    = {{USENIX} {OSDI}},
  publisher    = {{USENIX} Association},
  year         = {2026}
}

@inproceedings{DBLP:conf/fast/QinLHCRZ0ZX25,
  author       = {Ruoyu Qin and
                  Zheming Li and
                  Weiran He and
                  Jialei Cui and
                  Feng Ren and
                  Mingxing Zhang and
                  Yongwei Wu and
                  Weimin Zheng and
                  Xinran Xu},
  title        = {Mooncake: Trading More Storage for Less Computation - {A} KVCache-centric
                  Architecture for Serving {LLM} Chatbot},
  booktitle    = {{FAST}},
  pages        = {155--170},
  publisher    = {{USENIX} Association},
  year         = {2025}
}

@article{DBLP:journals/corr/abs-2510-09665,
  author       = {Yihua Cheng and
                  Yuhan Liu and
                  Jiayi Yao and
                  Yuwei An and
                  Xiaokun Chen and
                  Shaoting Feng and
                  Yuyang Huang and
                  Samuel Shen and
                  Kuntai Du and
                  Junchen Jiang},
  title        = {LMCache: An Efficient {KV} Cache Layer for Enterprise-Scale {LLM}
                  Inference},
  journal      = {CoRR},
  volume       = {abs/2510.09665},
  year         = {2025}
}

@inproceedings{DBLP:conf/sigcomm/LiuLCRHZDY0AMHH24,
  author       = {Yuhan Liu and
                  Hanchen Li and
                  Yihua Cheng and
                  Siddhant Ray and
                  Yuyang Huang and
                  Qizheng Zhang and
                  Kuntai Du and
                  Jiayi Yao and
                  Shan Lu and
                  Ganesh Ananthanarayanan and
                  Michael Maire and
                  Henry Hoffmann and
                  Ari Holtzman and
                  Junchen Jiang},
  title        = {CacheGen: {KV} Cache Compression and Streaming for Fast Large Language
                  Model Serving},
  booktitle    = {{SIGCOMM}},
  pages        = {38--56},
  publisher    = {{ACM}},
  year         = {2024}
}

@inproceedings{DBLP:conf/osdi/LeeLSS24,
  author       = {Wonbeom Lee and
                  Jungi Lee and
                  Junghwan Seo and
                  Jaewoong Sim},
  title        = {InfiniGen: Efficient Generative Inference of Large Language Models
                  with Dynamic {KV} Cache Management},
  booktitle    = {{OSDI}},
  pages        = {155--172},
  publisher    = {{USENIX} Association},
  year         = {2024}
}

@inproceedings{DBLP:conf/sigcomm/ZhuJJWSWZZWCXZL25,
  author       = {Ruidong Zhu and
                  Ziheng Jiang and
                  Chao Jin and
                  Peng Wu and
                  Cesar A. Stuardo and
                  Dongyang Wang and
                  Xinlei Zhang and
                  Huaping Zhou and
                  Haoran Wei and
                  Yang Cheng and
                  Jianzhe Xiao and
                  Xinyi Zhang and
                  Lingjun Liu and
                  Haibin Lin and
                  Li{-}Wen Chang and
                  Jianxi Ye and
                  Xiao Yu and
                  Xuanzhe Liu and
                  Xin Jin and
                  Xin Liu},
  title        = {MegaScale-Infer: Efficient Mixture-of-Experts Model Serving with Disaggregated
                  Expert Parallelism},
  booktitle    = {{SIGCOMM}},
  pages        = {592--608},
  publisher    = {{ACM}},
  year         = {2025}
}

@article{DBLP:journals/corr/abs-2507-19427,
  author       = {StepFun Inc.},
  title        = {Step-3 is Large yet Affordable: Model-system Co-design for Cost-effective
                  Decoding},
  journal      = {CoRR},
  volume       = {abs/2507.19427},
  year         = {2025}
}

@inproceedings{DBLP:conf/emnlp/AinslieLJZLS23,
  author       = {Joshua Ainslie and
                  James Lee{-}Thorp and
                  Michiel de Jong and
                  Yury Zemlyanskiy and
                  Federico Lebr{\'{o}}n and
                  Sumit Sanghai},
  editor       = {Houda Bouamor and
                  Juan Pino and
                  Kalika Bali},
  title        = {{GQA:} Training Generalized Multi-Query Transformer Models from Multi-Head
                  Checkpoints},
  booktitle    = {Proceedings of the 2023 Conference on Empirical Methods in Natural
                  Language Processing, {EMNLP} 2023, Singapore, December 6-10, 2023},
  pages        = {4895--4901},
  publisher    = {Association for Computational Linguistics},
  year         = {2023},
  url          = {https://doi.org/10.18653/v1/2023.emnlp-main.298},
  doi          = {10.18653/V1/2023.EMNLP-MAIN.298},
  bibsource    = {dblp computer science bibliography, https://dblp.org}
}

@article{DBLP:journals/corr/abs-1911-02150,
  author       = {Noam Shazeer},
  title        = {Fast Transformer Decoding: One Write-Head is All You Need},
  journal      = {CoRR},
  volume       = {abs/1911.02150},
  year         = {2019}
}

@article{DBLP:journals/corr/abs-2405-04434,
  author       = {DeepSeek{-}AI},
  title        = {DeepSeek-V2: {A} Strong, Economical, and Efficient Mixture-of-Experts
                  Language Model},
  journal      = {CoRR},
  volume       = {abs/2405.04434},
  year         = {2024},
  url          = {https://doi.org/10.48550/arXiv.2405.04434},
  doi          = {10.48550/ARXIV.2405.04434},
  eprinttype   = {arXiv},
  eprint       = {2405.04434},
  bibsource    = {dblp computer science bibliography, https://dblp.org}
}

@article{DBLP:journals/corr/abs-2401-04088,
  author       = {Albert Q. Jiang and
                  Alexandre Sablayrolles and
                  Antoine Roux and
                  Arthur Mensch and
                  Blanche Savary and
                  Chris Bamford and
                  Devendra Singh Chaplot and
                  Diego de Las Casas and
                  Emma Bou Hanna and
                  Florian Bressand and
                  Gianna Lengyel and
                  Guillaume Bour and
                  Guillaume Lample and
                  L{\'{e}}lio Renard Lavaud and
                  Lucile Saulnier and
                  Marie{-}Anne Lachaux and
                  Pierre Stock and
                  Sandeep Subramanian and
                  Sophia Yang and
                  Szymon Antoniak and
                  Teven Le Scao and
                  Th{\'{e}}ophile Gervet and
                  Thibaut Lavril and
                  Thomas Wang and
                  Timoth{\'{e}}e Lacroix and
                  William El Sayed},
  title        = {Mixtral of Experts},
  journal      = {CoRR},
  volume       = {abs/2401.04088},
  year         = {2024},
  url          = {https://doi.org/10.48550/arXiv.2401.04088},
  doi          = {10.48550/ARXIV.2401.04088},
  eprinttype   = {arXiv},
  eprint       = {2401.04088},
  bibsource    = {dblp computer science bibliography, https://dblp.org}
}

@inproceedings{DBLP:conf/acl/DaiDZXGCLZYWXLH24,
  author       = {Damai Dai and
                  Chengqi Deng and
                  Chenggang Zhao and
                  R. X. Xu and
                  Huazuo Gao and
                  Deli Chen and
                  Jiashi Li and
                  Wangding Zeng and
                  Xingkai Yu and
                  Y. Wu and
                  Zhenda Xie and
                  Y. K. Li and
                  Panpan Huang and
                  Fuli Luo and
                  Chong Ruan and
                  Zhifang Sui and
                  Wenfeng Liang},
  title        = {DeepSeekMoE: Towards Ultimate Expert Specialization in Mixture-of-Experts
                  Language Models},
  booktitle    = {{ACL} {(1)}},
  pages        = {1280--1297},
  publisher    = {Association for Computational Linguistics},
  year         = {2024}
}

@article{DBLP:journals/corr/abs-2505-09388,
  author       = {Qwen Team},
  title        = {Qwen3 Technical Report},
  journal      = {CoRR},
  volume       = {abs/2505.09388},
  year         = {2025}
}

@article{DBLP:journals/corr/abs-2412-19437,
  author       = {DeepSeek{-}AI},
  title        = {DeepSeek-V3 Technical Report},
  journal      = {CoRR},
  volume       = {abs/2412.19437},
  year         = {2024}
}

@inproceedings{DBLP:conf/iclr/LepikhinLXCFHKS21,
  author       = {Dmitry Lepikhin and
                  HyoukJoong Lee and
                  Yuanzhong Xu and
                  Dehao Chen and
                  Orhan Firat and
                  Yanping Huang and
                  Maxim Krikun and
                  Noam Shazeer and
                  Zhifeng Chen},
  title        = {GShard: Scaling Giant Models with Conditional Computation and Automatic
                  Sharding},
  booktitle    = {9th International Conference on Learning Representations, {ICLR} 2021,
                  Virtual Event, Austria, May 3-7, 2021},
  publisher    = {OpenReview.net},
  year         = {2021},
  url          = {https://openreview.net/forum?id=qrwe7XHTmYb},
  bibsource    = {dblp computer science bibliography, https://dblp.org}
}

@article{DBLP:journals/jmlr/FedusZS22,
  author       = {William Fedus and
                  Barret Zoph and
                  Noam Shazeer},
  title        = {Switch Transformers: Scaling to Trillion Parameter Models with Simple
                  and Efficient Sparsity},
  journal      = {J. Mach. Learn. Res.},
  volume       = {23},
  pages        = {120:1--120:39},
  year         = {2022},
  url          = {https://jmlr.org/papers/v23/21-0998.html},
  bibsource    = {dblp computer science bibliography, https://dblp.org}
}

@inproceedings{DBLP:conf/emnlp/BaiLZHQH0DL24,
  author       = {Yushi Bai and
                  Xin Lv and
                  Jiajie Zhang and
                  Yuze He and
                  Ji Qi and
                  Lei Hou and
                  Jie Tang and
                  Yuxiao Dong and
                  Juanzi Li},
  title        = {LongAlign: {A} Recipe for Long Context Alignment of Large Language
                  Models},
  booktitle    = {{EMNLP} (Findings)},
  series       = {Findings of {ACL}},
  pages        = {1376--1395},
  publisher    = {Association for Computational Linguistics},
  year         = {2024}
}

@inproceedings{dong2018tutorial,
  title={A tutorial on quantile estimation via Monte Carlo},
  author={Dong, Hui and Nakayama, Marvin K},
  booktitle={International Conference on Monte Carlo and Quasi-Monte Carlo Methods in Scientific Computing},
  pages={3--30},
  year={2018},
  organization={Springer}
}

@inproceedings{DBLP:conf/nips/DaoFERR22,
  author       = {Tri Dao and
                  Daniel Y. Fu and
                  Stefano Ermon and
                  Atri Rudra and
                  Christopher R{\'{e}}},
  title        = {FlashAttention: Fast and Memory-Efficient Exact Attention with IO-Awareness},
  booktitle    = {NeurIPS},
  year         = {2022}
}

@inproceedings{DBLP:conf/mlsys/00010LLZW0KGKC25,
  author       = {Zihao Ye and
                  Lequn Chen and
                  Ruihang Lai and
                  Wuwei Lin and
                  Yineng Zhang and
                  Stephanie Wang and
                  Tianqi Chen and
                  Baris Kasikci and
                  Vinod Grover and
                  Arvind Krishnamurthy and
                  Luis Ceze},
  title        = {FlashInfer: Efficient and Customizable Attention Engine for {LLM}
                  Inference Serving},
  booktitle    = {MLSys},
  publisher    = {OpenReview.net/mlsys.org},
  year         = {2025}
}

@misc{chatgpt,
  author       = {{OpenAI}},
  title        = {{ChatGPT}},
  howpublished = {\url{https://chatgpt.com}},
  year         = {2026},
  note         = {Accessed: 2026-05-08}
}

@misc{gemini,
  author       = {{Google}},
  title        = {{Gemini}},
  howpublished = {\url{https://gemini.google.com}},
  year         = {2026},
  note         = {Accessed: 2026-05-08}
}

@misc{deepseek,
  author       = {{DeepSeek-AI}},
  title        = {{DeepSeek}},
  howpublished = {\url{https://chat.deepseek.com}},
  year         = {2026},
  note         = {Accessed: 2026-05-08}
}

@misc{claudecode,
  author       = {{Anthropic}},
  title        = {{Claude Code}},
  howpublished = {\url{https://claude.com/product/claude-code}},
  year         = {2026},
  note         = {Accessed: 2026-05-08}
}

@misc{codex,
  author       = {{OpenAI}},
  title        = {{Codex}},
  howpublished = {\url{https://openai.com/codex}},
  year         = {2026},
  note         = {Accessed: 2026-05-08}
}

@misc{openclaw,
  author       = {{OpenClaw Contributors}},
  title        = {{OpenClaw}},
  howpublished = {\url{https://openclaw.ai}},
  year         = {2026},
  note         = {Accessed: 2026-05-08}
}

@misc{deepinfra,
  author       = {{DeepInfra}},
  title        = {{DeepInfra}},
  howpublished = {\url{https://deepinfra.com}},
  year         = {2026},
  note         = {Accessed: 2026-05-08}
}

@misc{alibabamodelstudio,
  author       = {{Alibaba Cloud}},
  title        = {{Alibaba Cloud Model Studio}},
  howpublished = {\url{https://modelstudio.alibabacloud.com}},
  year         = {2026},
  note         = {Accessed: 2026-05-08}
}

@misc{volcanoengine,
  author       = {{ByteDance}},
  title        = {{Volcano Engine}},
  howpublished = {\url{https://www.volcengine.com}},
  year         = {2026},
  note         = {Accessed: 2026-05-08}
}

@misc{nvidia-vmm,
  author       = {{NVIDIA}},
  title        = {{CUDA Virtual Memory Management (VMM)}},
  howpublished = {\url{https://docs.nvidia.com/cuda/cuda-driver-api/group__CUDA__VA.html}},
  year         = {2026},
  note         = {Accessed: 2026-05-08}
}

@misc{openrouter,
  author       = {{OpenRouter}},
  title        = {{OpenRouter}},
  howpublished = {\url{https://openrouter.ai}},
  year         = {2026},
  note         = {Accessed: 2026-05-08}
}

@misc{nvshmem,
  author       = {{NVIDIA}},
  title        = {{NVSHMEM: GPU Programming Interface for Scalable Communication}},
  howpublished = {\url{https://docs.nvidia.com/nvshmem}},
  year         = {2025},
  note         = {Accessed: 2026-05-08}
}

@misc{sharegpt-vicuna,
  title        = {Vicuna: An Open-Source Chatbot Impressing GPT-4 with 90\%* ChatGPT Quality},
  author       = {Wei-Lin Chiang and Lianmin Zheng and Ying Sheng and Tianle Li and others},
  year         = {2023},
  howpublished = {\url{https://lmsys.org/blog/2023-03-30-vicuna/}},
  note         = {Accessed: 2026-05-08}
}

@misc{sharegpt-vicuna-dataset,
  author       = {anon8231489123},
  title        = {ShareGPT Vicuna unfiltered},
  year         = {2023},
  howpublished = {\url{https://huggingface.co/datasets/anon8231489123/ShareGPT_Vicuna_unfiltered}},
  note         = {Accessed: 2026-05-08}
}

@misc{longalign10k,
  author       = {{Zhipu AI and Tsinghua University}},
  title        = {LongAlign-10k},
  howpublished = {\url{https://huggingface.co/datasets/zai-org/LongAlign-10k}},
  year         = {2024}
}

@misc{deepseek-v4,
  title        = {DeepSeek-V4: Towards Highly Efficient Million-Token Context Intelligence},
  author       = {{DeepSeek-AI}},
  year         = {2026},
  howpublished = {\url{https://huggingface.co/deepseek-ai/DeepSeek-V4-Pro/blob/main/DeepSeek_V4.pdf}},
  note         = {Technical Report}
}

\balance

\end{document}